\documentclass[12pt,letterpage]{amsart}
\usepackage{amsmath}
\newcommand{\be}{\begin{equation}}
\newcommand{\ee}{\end{equation}}

\usepackage{color}
\definecolor{orange}{rgb}{1,0.5,0}
\usepackage{graphics}
\usepackage{wrapfig}
\usepackage{amsmath}
\usepackage{amssymb}
\usepackage{subfigure}
\usepackage{float}
\usepackage{floatflt}
\usepackage{hyperref}
\usepackage{amssymb,amsmath}
\usepackage{graphicx}
\usepackage{epstopdf}
\usepackage{fancyhdr}
\begin{document}

\title{Single Layer Valves in Microfluidics} 
\author{Natalie Arkus}
\address{Center for Studies in Physics and Biology,
The Rockefeller University, New York, NY 10065 }

\begin{abstract}
A mechanism for constructing single layer valves in microfluidic devices is reported.
\end{abstract}

\maketitle

Here we report the fabrication of single layer valves in microfluidics.  This mechanism of valve construction was created several years ago, but was never optimized.  We nonetheless felt it useful to report this mechanism so that others could use it, and perhaps even optimize it.

The ability to close channels in microfluidic devices and to re-open and re-close them on demand is of great use in experiments.  There is already a well-known and well-used microfluidic valve mechanism for doing this \cite{WhitesidesValves,QuakeValves,StanfordFoundryWebsiteValves}.  However, multiple microfluidic layers are used to construct these valves, and multilayer devices are significantly more difficult to construct than single layer ones, greatly reducing the ease of microfluidic fabrication.  Here, we adopt the existing multilayer valve technology to single layers, so that one can have the utility of valves along with the ease of single layer fabrication.
%Make first paragraph shorter...more overview, leading into going into it more...
%background of what kinds of experiments these are used in
%background of quake's method
%How we adopt it
%pictures
%roadfalls we came into
%other people adopting it
%conclusion -- that it's possible and easier, but needs to be troubleshot...

%\begin{figure}[htbp]
%\begin{center}
%%\includegraphics[width = 0.3\textwidth]{Wafer2_Chip17_crop3}
%\includegraphics[width = 0.3\textwidth]{Wafer2_Chip17_cropOneSide2}
%\includegraphics[width = 0.3\textwidth]{ValvesMaskCrop3}
%%\includegraphics[width = 0.3\textwidth]{MaskZoomIn1}
%\includegraphics[width = 0.3\textwidth]{ValvesMaskPic2}
%\caption{default}
%\label{default}
%\end{center}
%\end\figure}

\begin{figure}[h!]
\begin{center}
\includegraphics[width = 1.0\textwidth]{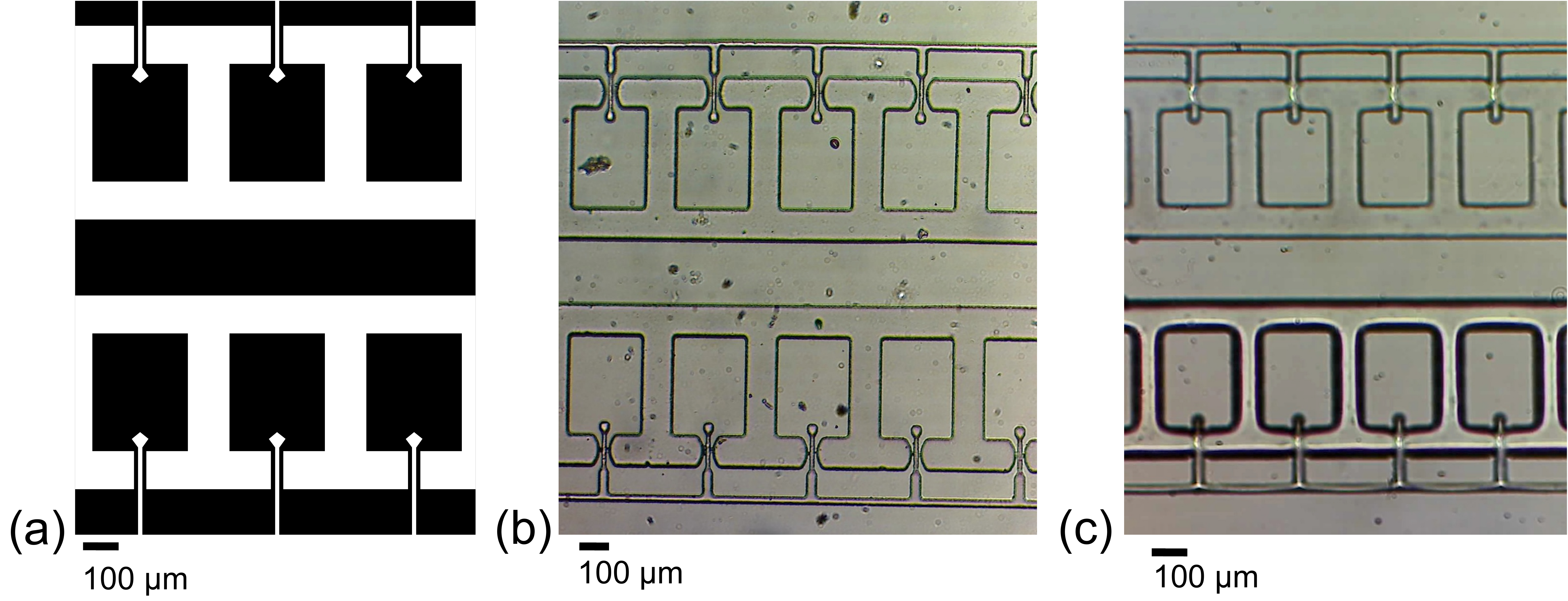}
\caption{Microfluidic chips containing single layer valves.  Each chip contains identical top and bottom sections, mirror images of one another.  Unless otherwise stated, each section contains a $100\mu$m main channel with T-shaped valves emanating out of it, and a $20\mu$m main channel with wells shooting off it; with offshoots the same width as the main channel.  Chips are approximately $20\mu$m in height.  The top of the T valve is $100\mu$m x $300\mu$m.  There is a $5\mu$m separation between the valves and the well channels.  (a) Portion of a mask used to create one of the chips containing square single layer valves.  Channels are white.  Note the wells are round, $40\mu$m in diameter, but appear diamond shaped as an artifact of the pdf viewer.  The main well channel (not shown) is $100\mu$m in width, and its offshooting channels are $10\mu$m.  (b) A chip with rounded valves.  The well channel offshoots narrow before opening into a $20\mu$m square well, creating a distinct well even when the valves are open.  c) A chip with rounded valves and straight well channel offshoots.  The wells in this chip are created only when the valves are closed.  The valves in the bottom section of the chip are inflated and encroach in on the well channels between them.  However, note this is a defective chip, with valve and well channels bleeding into one another -- it is only shown here as an illustration of inflated valves and channels and could not be used for proper valve closures.}
\label{ValvePic}
\end{center}
\end{figure}

Microfluidic fabrication involves coating a silicon wafer with photoresist, overlaying a mask containing the device's pattern, and exposing it to ultraviolet (uv) light.  The uv cross-links the photoresist and forms a mold of the device.  The microfluidic devices, or chips, are then made by pouring a soft polymer, such as polydimethylsiloxane (PDMS), over the master mold, and baking it with an elastomeric curing agent that cross-links the polymer, turning it into a flexible solid.  The chips are then sealed to a glass slide.  One master can be used to make many devices.\cite{XiaWhitesides, WhitesidesStandard}  Note, here on in, we will refer to PDMS devices, but generally speaking any appropriate soft polymer can be used.  The applications of microfluidic devices are vast; from the study of colloidal systems \cite{Meng29012010, ChaikinPineColloid, WeitzColloid}, to fluid dynamics \cite{Stone, RevModPhys.77.977}, to molecular biology \cite{electrophoresis, ZhangPCR}, to cell culture studies \cite{BactChemoMicro, QuakeCellCulture, Tay2010}, even to evolutionary and population biology \cite{Keymer14112006, Jordan20082013}, and this is only to name a few.  Part of the widespread applicability of microfluidics is due to its relative ease of fabrication.  Learning the basics of microfluidic fabrication can allow one to design a novel device and to construct it from scratch in less than a week (the most time consuming step not having to do with fabrication, but with waiting for the printed mask to be shipped).  And with facilities such as the Stanford Foundry, one no longer needs to have access to a clean room or to know how to make their own master or even their own device, as both can be ordered.

The existing microfluidic valve technology involves a device with at least two layers; a flow and a control layer \cite{WhitesidesValves,QuakeValves,StanfordFoundryWebsiteValves}.  The flow layer contains channels through which the fluids flow and relevant experiments can take place, and the control layer contains channels that deflect into the flow channels and seal them off when they are pressurized with either air or liquid -- these are the valves.  Valve and flow channels are typically positioned perpendicular to one another.  The fidelity of the device depends upon the precision of the alignment of the layers.  One mold is made for the control layer and one mold is made for the flow layer.  The PDMS casts of each layer are then aligned.  The resulting valves can be closed and opened through inflation/deflation of the control channels repeatedly without delimanitation.  While it is common practice to align layers and create microfluidic devices with valves, it is this alignment step that is both tedious and time-consuming when compared to the ease of constructing a single layer cast.  The more microfluidic chips one requires for an experiment, the more severe this difference is felt as the layers of each chip must be aligned by hand under a microscope.  It would be a great advantage to be able to construct microfluidic valves with the ease of single layer devices; this would not only facilitate single chip fabrication but would also make high throughput experiments using these devices more practical.

%good, but perhaps streamline, and perhaps repetitive with earlier - so make unrepetitive...
The multilayer valve technology makes use of the deformability of the soft polymer to create an inflated control channel that, along with the surrounding polymer, expands into the flow channel, thereby sealing it.  This deformability is reversible; once the control channels are emptied, they deflate, the surrounding polymer rebounds, and nothing encroaches on the flow channel, thereby opening it.  This basic process is not particular to multiple layers, and the act of reversible polymer deformation creating one channel that encroaches in on/pulls away from another when inflated/deflated should be realizable horizontally in addition to vertically, and thus should be constructible in a single layer.  

%also streamline with what comes earlier
Here, we create flow and control channels within the same layer.  As with the multilayer devices, the flow and control channels are separated by a thin layer of PDMS -- the difference here being that the PDMS lies horizontally between two side-by-side channels; whereas, in multilayer devices the PDMS lies vertically between two channels, one on top of the other.  The control channels contain one inlet, through which air or fluid can be supplied to pressurize the channel, and no outlet.  Once the channel is pressurized, it can remain so for a prolonged amount of time, allowing for long term closures of fluid channels, if desired.  The control channel is deflated when the syringe pulls the fluid or air back out, or when the syringe is removed all together.  The flow channels contain both an inlet and an outlet.

Figure \ref{ValvePic} shows an implementation of single layer valves.  These are T-shaped valves, where there is a main channel, with T-shaped offshoots.  These chips contain identical top and bottom sections, mirror images of one another.  The flow channel consists of one main channel with offshooting well channels.  The well, contained at the end of the offshoot is either formed via the valve closing, or via a narrowing of the flow channel leading into the well.  In the latter case, the well is distinctly formed even when the valves are open; the closing of the valves seals the wells so that they are isolated.  The purpose of these chips was to dispense colloidal particles or cells into the wells, and to seal the wells, at which point the colloidal system could reach equilibrium or the cells could be observed under a given condition.  Top and bottom well/valve sections each contain on the order of 30 wells and valves.  Valve channels are $100\mu$m in width, with the top of the T being $300\mu$m x $100\mu$m.  Unless otherwise specified, wells are $20\mu$m square or $20\mu$m in diameter for square and circular wells, respectively.  We explored a narrow range of separation lengths, but there is typically a 5 or $10\mu$m PDMS separation between flow and valve channels on either side of the top of the T.  The channels are approximately $20\mu$m high.  We explored a concentration of 5-15\% curing agent, all of which worked comparably, and settled on using the 5\% concentration.

The top of the T is where the valves close.  Each well channel is closed by two T-valves, half of each T inflating into the flow channel from either side (fig \ref{ValvePic}).  In these devices, when the control channel is pressurized, all of the T-valves are inflated, sealing all the wells off at once.  However, devices can easily be constructed in which separate valve channels close different wells at different times.  In a functioning chip, the T-valves inflate to the point of appearing to touch under up to 60x magnification of an optical microscope (data not shown).  The entry into the well appears completely pinched off between the inflated T-valves during the closure.  When the control channels are depressurized, flow and valve channels return to their normal state, and this has been repeated many times without delamination (see movie).  We have explored different flow and control channel topologies and dimensions, and the fidelity of the valve closure appears to be insensitive to factors such as the roundedness or squareness of the valves.  Each valve topology that we have examined involves two valve channels sandwiching the flow channel.  While we have explored several different flow and valve dimensions, they do not differ significantly from the ones described here.  Colleagues have adopted this single layer valve concept and have fabricated a chip in which the flow channel is not sandwiched between 2 valves, but rather is closed by a single protruding valve.  In this case, both flow and control channels were much larger (greater than $100\mu$m), and air pressure was supplied via a gas outlet at a pressure estimated around 20psi as opposed to manually controlled syringes, as with the devices described here.  The valves did close the flow channel, but the seal was not complete. \cite{SeppeCommunication}

%again, and until end, word better with rest, and also, have to add in some more info on dimensions, etc....
While we have created functioning single layer valve chips, we have also encountered many problems.  While we have created some chips whose valves can be open and closed repeatedly (see movie), we have often run into problems with delamination.  And while we have created some chips in which the flow channels are completely sealed off by the valves, we have created many chips in which the valves do not seal all the way, or even fail to encroach on the flow channels all together.  These problems have arisen across different chip designs, but also within different implementations of the same chip design.  We have not determined the source of these problems; however, we suspect it may have to do with the fidelity of the sealing of the chip to the glass slide.

In summary, we have created functional single layer microfluidic valves; however, we have not been able to construct them reliably.  Reliable construction of single layer microfluidic valves would allow for high-throughput experiments with sealed sub-environments.  We report this mechanism for constructing single layer valves in the hopes that others might find it useful, or even be able to optimize it, yielding an easy and reliable single layer microfluidic valve technology.

\textit{Materials and  Methods:} Microfluidic devices were fabricated using soft lithography.  Photomask patterns were designed in AutoCAD, QCAD, or TurboCAD.  Master molds were either fabricated by the author in a clean room or by the Stanford Foundry (http://www.stanford.edu/group/foundry/).  For devices fabricated by the author, the photomask pattern was printed on a photomask transparency by CAD/Art Services, Inc. (http://www.outputcity.com/).  Master molds were then made by spinning SU-8 photoresist (SU-8 2025 or 2075, MicroChem Corp.) onto a silicon wafer (University Wafer) in a clean room environment using standard procedures in photolithography \cite{WhitesidesStandard}.  Microfluidic devices were then fabricated by pouring a polydimethylsiloxane (PDMS) cross-linker mixture (Sylgard 184 silicone elastomer and curing agent, Dow Corning Corporation, Midland, MI, USA) over the master and curing it in the oven for 1-2 hours at 80$^\circ$C.  We used a 5\% concentration of the curing agent.  Chips were then sealed to glass slides using plasma bonding.
\\
\\

We thank Stanislas Leibler for his generosity in allowing us to use his clean room, Seppe Kuehn and David Jordan for helpful consultations in clean room usage and microfluidic fabrication, Timothy Halpin-Healy for his help in providing access to the Columbia University clean room, Howard Stone and Jiandi Wan for providing initial training in microfluidic fabrication and clean room usage, John Lee for technical help with software, and Benjamin Greenbaum for helpful discussions.

\bibliographystyle{ieeetr}
\bibliography{SingleLayerValvesBib}

\end{document}